\documentclass[a4paper,fleqn,usenatbib]{mn2e}

\usepackage[T1]{fontenc}
\usepackage{ae,aecompl}

\usepackage{graphicx}	

\def\la{\;
\raise0.3ex\hbox{$<$\kern-0.75em\raise-1.1ex\hbox{$\sim$}}\; }
\def\ga{\;
\raise0.3ex\hbox{$>$\kern-0.75em\raise-1.1ex\hbox{$\sim$}}\; }

\newcommand{\daa}{$\Delta\alpha/\alpha$}
\newcommand{\dmm}{$\Delta\mu/\mu$}
\newcommand{\kms}{km~s$^{-1}$}
\newcommand{\etal}{{et al.}}

\title[ {\rm [C\,{\sc i}]}, {\rm [C\,{\sc ii}]} and {\rm CO} emission lines as 
a probe for $\alpha$ variations]{{\rm\bf [C\,{\sc i}]}, 
{\rm\bf [C\,{\sc ii}]} and {\rm\bf CO} emission lines as a probe for {\bf $\alpha$} variations at low and high redshifts}

\author[S. A. Levshakov et al.]{
S. A. Levshakov,$^{1,2,3}$\thanks{E-mail: lev@astro.ioffe.ru}
K.-W. Ng,$^{4,5}$
C. Henkel$^{6,7}$
and B. Mookerjea$^{8}$
\\
$^{1}$Ioffe Physical-Technical Institute, 194021 St.~Petersburg, Russia\\
$^{2}$St.~Petersburg Electrotechnical University ``LETI'', 197376 St.~Petersburg, Russia\\
$^{3}$ITMO University, 191002 St.~Petersburg, Russia\\
$^{4}$Institute of Physics, Academia Sinica, Taipei 11529, Taiwan\\
$^{5}$Institute of Astronomy and Astrophysics, Academia Sinica, Taipei 11529, Taiwan\\
$^{6}$Max Planck Institut f\"ur Radioastronomie, Auf dem H\"ugel 69, 53121 Bonn, Germany\\
$^{7}$Astron. Dept., King Abdulaziz University, PO Box 80203, 21589 Jeddah, Saudi Arabia\\
$^{8}$Tata Institute of Fundamental Research, Homi Bhabha Road, 400005 Mumbai, India
}

\date{Accepted XXX. Received YYY; in original form ZZZ}

\pubyear{2017}

\begin{document}
\label{firstpage}
\pagerange{\pageref{firstpage}--\pageref{lastpage}}
\maketitle

\begin{abstract}
The offsets between the radial velocities of the
rotational transitions of carbon monoxide and
the fine structure transitions of neutral and singly ionized carbon
are used to test the hypothetical variation of the fine structure constant, $\alpha$.
From the analysis of the [C\,{\sc i}] and [C\,{\sc ii}] fine structure lines and low $J$
rotational lines of $^{12}$CO and $^{13}$CO, emitted by 
the dark cloud L1599B in the Milky Way disk,
we find no evidence for fractional changes in $\alpha$ at the level of
$|\Delta\alpha/\alpha| < 3\times10^{-7}$.
For the neighbour galaxy M33 a stringent limit on \daa\
is set from observations of three H\,{\sc ii} zones in [C\,{\sc ii}]
and CO emission lines: $|\Delta\alpha/\alpha| < 4\times10^{-7}$.
Five systems over the redshift interval $z = 5.7-6.4$, 
showing CO $J=6-5$, $J=7-6$ and [C\,{\sc ii}] 158 $\mu$m emission,
yield a limit on $|\Delta\alpha/\alpha| < 1.3\times10^{-5}$.
Thus, a combination of the [C\,{\sc i}], [C\,{\sc ii}], and CO emission
lines turns out to be a powerful tool for probing the stability of the fundamental
physical constants over a wide range of redshifts not accessible to
optical spectral measurements.
\end{abstract}


\section{Introduction}

The variability of dimensionless physical constants
such as the electron-to-proton mass ratio, $\mu = m_{\rm e}/m_{\rm p}$,
and the fine structure constant, $\alpha = e^2/\hbar c$, 
remains an active area of theoretical and experimental studies 
(e.g., Bainbridge \etal\ 2017; Levshakov \& Kozlov 2017; Martins \& Pinho 2017; Thompson 2017; 
Hojjati \etal\ 2016; Levshakov 2016).
In the Standard Model (SM) of particle physics, $\alpha$ and $\mu$
are not supposed to vary in space and time.
Any such variation would be a violation of the Einstein Equivalence
Principle (EEP) and an indication of physics beyond the SM 
(e.g., Liberati 2013; Uzan 2011; Kosteleck\'y \etal\ 2003).
Thus, laboratory and astrophysical studies aimed at measuring 
the relative deviation in the fine structure constant and/or in the electron-to-proton mass ratio
from the current value, \daa\ and/or \dmm,
are the most important tools to test new theories against 
observations\footnote{$\Delta\alpha/\alpha = (\alpha_{\rm obs} - \alpha_{\rm lab})/\alpha_{\rm lab}$,
$\Delta\mu/\mu = (\mu_{\rm obs} - \mu_{\rm lab})/\mu_{\rm lab}$.}.

The validity of EEP and general relativity was recently confirmed at high precision
both in laboratory experiments with atomic clocks and in astrophysical 
observations\footnote{Below and throughout the paper 1$\sigma$ values are provided.}. 
The atomic clock measurements at
time scales of the order of one year restrict the temporal rate of change in $\alpha$
at the level of
$\dot{\alpha}/\alpha < 10^{-17}$ yr$^{-1}$ 
(Godun \etal\ 2014; Leefer \etal\ 2013; Rosenband \etal\ 2008).
For a shorter period ($t \la 1000$ s), 
a fractional frequency uncertainty in atomic clocks
is $\Delta\omega/\omega \sim 10^{-18}$
(Huntemann \etal\ 2016; Nisbet-Jones \etal\ 2016; Nicholson \etal\ 2015),
which can provide even tighter constraints on $\dot{\alpha}/\alpha$ 
with longer time intervals\footnote{The
relationship between $\Delta\omega/\omega$ and $\Delta\alpha/\alpha$ is
given in, e.g., Kozlov \& Levshakov 2013.}.

In agreement with these laboratory experiments,
the stability of $\alpha$ over the past $2\times10^9$ yr follows
from the Oklo phenomenon~-- the uranium mine in Gabon, 
which provides the tightest terrestrial bound on $\alpha$ variation:
$\dot{\alpha}/\alpha < 5\times10^{-18}$ yr$^{-1}$ (Davis \& Hamdan 2015).
We note that the age $2\times10^9$ yr corresponds to the cosmological epoch
at redshift $z = 0.16$ if a cosmology with $H_0 = 70$ km~s$^{-1}$~Mpc$^{-1}$,
$\Omega_\Lambda = 0.7$, and $\Omega_m = 0.3$ is adopted. 

In astrophysical tests, the EEP was validated over the redshift range $z = 0.03 - 0.9$
from measurements of
time delays between different energy bands observed from blazar flares,
gamma-ray bursts, and fast radio bursts 
(Petitjean \etal\ 2016; Wei \etal\ 2016; Gao \etal\ 2015).
These experiments set limits on the numerical coefficients of 
the parameterized post-Newtonian formalism, in particular, 
on the parameter $\gamma$ accounting for how much space-curvature is produced by
unit rest mass. The general relativity predicts $\gamma \equiv 1$, which was confirmed
at a level of $\sim 10^{-8}$.

Traditional spectral line observations of Galactic and extragalactic
objects provide us with independent constraints.
Thus, from metal absorption-line spectra of quasars the variation in $\alpha$ was 
limited at a few $10^{-6}$ in the redshift range $z \sim 1-3$
(Kotu$\check {\rm s}$ \etal\ 2017; Murphy \etal\ 2017; 
Evans \etal\ 2014; Molaro \etal\ 2013; Agafonova \etal\ 2011;
Levshakov \etal\ 2006; Quast \etal\ 2004).

From optical observations of the Lyman and Werner systems of H$_2$ and HD 
and the forth positive system of CO in quasar absorption-line spectra 
the fractional changes in the electron-to-proton mass ratio \dmm\
were determined to an accuracy of $8\times10^{-6}$ at $z = 4.2$ (Bagdonaite \etal\ 2015)
and to the same order of magnitude over the redshift range $z = 2.3-3.0$ 
(Dapr\`a \etal\ 2017b,c; Vasquez \etal\ 2014; Rahmani \etal\ 2013;
Wendt \& Molaro 2012; King \etal\ 2011; King \etal\ 2008; Wendt \& Reimers 2008).

A more stringent limit was obtained for \dmm\ from radio observations 
of molecular absorption lines of NH$_3$, CS, H$_2$CO, and CH$_3$OH which lead to
the $1\sigma$ constraint
$|\Delta\mu/\mu| < (1-2)\times10^{-7}$ at $z \sim 0.7-0.9$ 
(Marshall \etal\ 2017; Kanekar \etal\ 2015; Bagdonaite \etal\ 2013a,b;
Ellingsen \etal\ 2012; Kanekar 2011; Henkel \etal\ 2009).

High-resolution spectral observations of molecular emission lines of NH$_3$,
HC$_3$N, HC$_5$N, HC$_7$N, and several CH$_3$OH transitions towards dark clouds
in the Milky Way disk are consistent with zero \dmm\ at an order of magnitude deeper 
level of
$0.7\times10^{-8}$ (Levshakov \etal\ 2013) and $(2-3)\times10^{-8}$ 
(Dapr\`a \etal\ 2017a; Levshakov \etal\ 2011).
These new results improve earlier constraints on \dmm\ based on spectral observations
of molecular absorption lines of H$_2$, HD, and CO in high-redshift intervening systems 
by almost three orders of magnitude and are competitive with 
precise laboratory measurements (see a review by Kozlov \& Levshakov 2013). 
The most precise constraints on variations of the mentioned above physical constants
are summarized in Table ~\ref{T0}. 

Since radio observations provide an essential progress in testing \dmm\
at a deeper level as compared with optical measurements,
a question arises as to whether the non-variability of $\alpha$ in space and time can be proved 
in the radio sector with a better precision than $10^{-6}$~--- the optical limit. 
We note that the fine structure constant is expected to vary at most at the $10^{-7}$ level 
within the framework of models of modified gravity of the chameleon type (e.g., Brax 2014)
or $f(T)$ gravity models (e.g., Nunes \etal\ 2017),
but corresponding observational constraints are still missing.

The improvement of the optical limit on \daa\ by an order of magnitude seems to be possible
through the analysis of far infrared (FIR) and sub-mm lines observed in extragalactic objects.
For instance, the fine structure (FS) 
[C\,{\sc ii}] 158 $\mu$m line~--- one of the main cooling agents of the 
interstellar medium (ISM)~--- is widely observed in distant galaxies (e.g., Hemmati \etal\ 2017).
The sensitivity of this FS line to small changes in $\alpha$ 
is about 30 times higher than that of ultra-violet (UV) lines of atoms and ions employed in optical
spectroscopy (Kozlov \etal\ 2008).
Weaker, but also prominent are the FS lines of neutral carbon, [C\,{\sc i}]~609, 370 $\mu$m.
A comparison between the observed frequencies of
the [C\,{\sc ii}] or [C\,{\sc i}] FS lines and the pure rotational CO transitions
is sensitive to changes in the ratio $F = \alpha^2/\mu$ (Levshakov \etal\ 2008):
\begin{equation}
\frac{\Delta F}{F} \equiv \frac{\Delta V}{c} = 2\frac{\Delta \alpha}{\alpha} - \frac{\Delta \mu}{\mu}\ ,
\label{Eq1}
\end{equation}
where $\Delta V = V_{\rm rot} - V_{\rm fs}$ is the difference between the radial velocities of the CO
rotational line and the [C\,{\sc ii}] or [C\,{\sc i}] FS lines. $c$ is the speed of light.

The velocity offset $\Delta V$ in this equation can be represented by the sum of two components:
\begin{equation}
\Delta V = \Delta V_F + \Delta V_D\ ,
\label{Eq1a}
\end{equation}
where $\Delta V_F$ is the shift due to $F$-variation, whereas $\Delta V_D$ is the Doppler noise,
which is a random component caused by local effects,
since different tracers may arise
from different parts of gas clouds, at different radial velocities.
The Doppler noise may either mimic or obliterate a real signal. 
However, if these offsets are random, the signal $\Delta V_F$
can be estimated statistically by averaging over a data sample:
\begin{equation}
\langle \Delta V \rangle = \langle \Delta V_F \rangle , \,\,\, 
{\rm Var}(\Delta V) = {\rm Var}(\Delta V_F) + {\rm Var}(\Delta V_D),
\label{Eq1b}
\end{equation}
where the noise component is assumed to have a zero mean and a finite variance (Levshakov \etal\ 2010).

This technique was used to analyze the $\alpha^2/\mu$ ratio 
for two high-redshift [C\,{\sc ii}]/CO systems at $z = 6.42$ with $\Delta F/F = (0.1 \pm 1.0)\times10^{-4}$, 
and at $z = 4.69$ with $\Delta F/F = (1.4 \pm 1.5)\times10^{-4}$ (Levshakov \etal\ 2008),
five Galactic molecular clouds with 
$\Delta F/F = (0.7 \pm 2.7_{\rm stat} \pm 1.1_{\rm sys})\times10^{-7}$ (Levshakov, Molaro \& Reimers 2010),
eight [C\,{\sc i}]/CO systems over $z = 2.3 - 4.1$ with $\Delta F/F = (-3.6 \pm 8.5)\times10^{-5}$ 
(Curran \etal\ 2011),
one [C\,{\sc i}]/CO system at $z = 5.2$ with $\Delta F/F = (0.3 \pm 1.5)\times10^{-5}$ (Levshakov \etal\ 2012), 
and one 
[C\,{\sc i}]/CO system at $z = 2.79$ with $\Delta F/F = (6.9 \pm 3.7)\times10^{-6}$ (Wei\ss\ \etal\ 2012).

Theory predicts that
the variation of $\mu$ should be larger than that of $\alpha$ 
(e.g. Langacker \etal\ 2002; Flambaum 2007),
but observational confirmation of this suggestion is still missing, so that
potential variations in $\alpha$ still need to be checked.
Taking into account that 
the limit on \dmm\ is about $10^{-8}$ in Galactic dark clouds
and that dark clouds in nearby galaxies exhibit the same physical properties
as those in the Milky Way disk (i.e., gas densities, kinetic and excitation temperatures, masses etc.),
we may suppose that the results on \dmm\ 
from Galactic dark clouds hold for the entire local Universe and
neglect the second term in the right hand side of Eq.(\ref{Eq1}) 
re-writing this equation in the form: 
\begin{equation}
\frac{\Delta \alpha}{\alpha} = \frac{\Delta V}{2c}\ .
\label{Eq2}
\end{equation}

As long as the neglected term imposes an error
of the order of $10^{-8}$, this approximation
is justified in the interval $10^{-7} \la |\Delta\alpha/\alpha| \la 10^{-6}$ 
of putative variations of $\alpha$.

In this approach, CO rotational transitions (being independent on $\alpha$)
serve as an anchor and reference position of the radial velocity scale,
and the variation in $\alpha$ manifests as a velocity offset between the observed
positions of the fine structure and CO transitions when compared to rest frame frequencies.
The described procedure is an analog to the Fine Structure Transition (FST) method proposed in
Levshakov \& Kozlov (2017).

Here, we will apply this method to the [C\,{\sc i}], [C\,{\sc ii}] and CO emission lines,
observed in nearby galaxies with orbital observatories and ground-based radio telescopes.
We also update our previous results on molecular clouds in the Milky Way
and on quasar systems at redshifts $z > 5.7$.

\begin{table}
\centering
\caption{The most precise differential measurements 
of the dimensionless physical
constants $\alpha$, $\mu$, and $\gamma$ ($1 \sigma$ upper bounds). 
Listed are, respectively, the method used,
the parameter being constrained and the time or redshift interval, and the original references.
The time and the redshift scales are shown in Fig.~\ref{Fg3}.
}
\label{T0}
\begin{tabular}{lll} 
\hline
{\tiny atomic clocks} & {\small $\dot{\alpha}/\alpha < 10^{-17}$ yr$^{-1}$} & {\tiny Godun et al. 2014}\\[-4pt] 
 &  & {\tiny Leefer et al. 2013}\\[-4pt] 
 &  & {\tiny Rosenband et al. 2008}\\ 
{\tiny Oklo phenomenon} & {\small $\dot{\alpha}/\alpha < 5\times10^{-18}$ yr$^{-1}$} & {\tiny Davis \& Hamdan 2015}\\
{\tiny $\gamma$ parameter in} & {\small $\Delta\gamma/\gamma < 10^{-8}$} & {\tiny Petitjean et al. 2016}\\[-4pt] 
{\tiny post-Newtonian } & {\tiny at $z=0.03-0.9$} & {\tiny Wei et al. 2016}\\[-4pt] 
{\tiny models} & {\tiny ($\gamma \equiv 1$ in general relativity)} & {\tiny Gao et al. 2015}\\
{\tiny quasar} & {\small $\Delta\alpha/\alpha <$ a few $10^{-6}$} & {\tiny Kotu$\check {\rm s}$ et al. 2017}\\[-4pt] 
{\tiny absorption-line} & {\tiny at $z \sim 1-3$ } & {\tiny Murphy et al 2017}\\[-4pt]
{\tiny spectra } & & {\tiny Evans et al. 2014}\\[-4pt]
 & & {\tiny Molaro et al. 2013}\\[-4pt] 
 & & {\tiny Agafonova et al. 2011}\\[-4pt]
 & & {\tiny Levshakov et al. 2006}\\[-4pt] 
 & & {\tiny Quast et al. 2004}\\
{\tiny H$_2$, HD, CO} & {\small $\Delta\mu/\mu <$ a few $10^{-6}$} & {\tiny Bagdonaite et al. 2015}\\[-4pt]
{\tiny absorption } & {\tiny at $z=4.2$ } & {\tiny Dapr\`a et al. 2017b,c}\\[-4pt] 
{\tiny lines in }   & {\tiny and at $z \sim 2-3$ } & {\tiny Vasquez et al. 2014}\\[-4pt] 
{\tiny quasar } & & {\tiny Rahmani et al. 2013}\\[-4pt]
{\tiny spectra} & & {\tiny Wendt \& Molaro 2012}\\[-4pt] 
 & & {\tiny King et al. 2011, 2008}\\[-4pt] 
 & & {\tiny Wendt \& Reimers 2008}\\
{\tiny NH$_3$, CS, H$_2$CO, } & {\small $\Delta\mu/\mu < (1-2)\times10^{-7}$} & {\tiny Marshall et al. 2017}\\[-4pt]
{\tiny CH$_3$OH } & {\tiny at $z \sim 0.7-0.9$} & {\tiny Kanekar et al. 2015}\\[-4pt] 
{\tiny absorption } & & {\tiny Bagdonaite et al. 2013a,b}\\[-4pt]
{\tiny lines in} & & {\tiny Ellingsen et al. 2012}\\[-4pt] 
{\tiny galaxies} & & {\tiny Kanekar 2011}\\[-4pt]
 & & {\tiny Henkel et al. 2009}\\
{\tiny NH$_3$, HC$_3$N,} & {\small $\Delta\mu/\mu < 0.7\times10^{-8}$} & {\tiny Levshakov et al. 2013}\\[-4pt]
{\tiny HC$_5$N, HC$_7$N,} & {\tiny and}  & {\tiny Dapr\`a et al. 2017a}\\[-4pt]
{\tiny CH$_3$OH }        & {\small $\Delta\mu/\mu < (2-3)\times10^{-8}$} &  {\tiny Levshakov et al. 2011}\\[-4pt]
{\tiny emission lines} & {\tiny at $z=0$ }\\[-4pt]
{\tiny in dark clouds of} & & \\[-4pt]
{\tiny the Galactic disk} & &\\
\hline
\end{tabular}
\end{table}

\section{Analysis and results}

The major cooling lines of the ISM in galaxies~--- [C\,{\sc i}], [C\,{\sc ii}], and [O\,{\sc i}]~---
trace the transition regions between the atomic and molecular gas.
[C\,{\sc ii}] emission can arise 
in the diffuse warm ionized medium, H\,{\sc ii} regions, and mostly neutral
photon dominated regions (PDRs)
because of the low ionization potential of carbon
(11.26 eV versus 13.60 eV for hydrogen). 
Observations with high angular resolution show similarity in shape
between the [C\,{\sc ii}] and CO profiles (e.g., de Blok \etal\ 2016; Braine \etal\ 2012).
This allows us to compare the radial velocities of the [C\,{\sc ii}] FS line and the
CO rotational lines
at different positions within a given galaxy
and to average a sample of the offsets $\Delta V = V_{\rm rot} - V_{\rm fs}$ to minimize 
the Doppler noise~---
random shifts of the measured spectral positions caused by deviations from co-occurrence in
space between these species.

The rest frame frequency of the [C\,{\sc ii}] $^2$P$_{3/2}$--$^2$P$_{1/2}$ line is
$\omega = 1900536.9$ MHz, known to within an uncertainty of $\sigma_\omega = 1.3$ MHz, i.e.
$\sigma_v$([C\,{\sc ii}]) = 0.2 km~s$^{-1}$ (Cooksy \etal\ 1986a), 
whereas the rest frame frequency of, for example, the CO $J=2$--1 line
is $\omega = 230538.000$ MHz 
and its uncertainty is only $\sigma_\omega = 0.5$ kHz,
i.e. $\sigma_v$(CO(2-1)) = 0.007 km~s$^{-1}$ (Endres \etal\ 2016).
Therefore, the utmost sensitivity of the FST method for the combination of the [C\,{\sc ii}] FS and CO(2-1)
transitions is mainly restricted by
the quoted error $\sigma_v$[C\,{\sc ii}] and equals to  
$\sigma^{lim}_{\Delta\alpha/\alpha} \simeq 3\times10^{-7}$.

To collect a sample of $\Delta V$ offsets, we apply the following selection criteria: 
($i$) the [C\,{\sc ii}] and CO emission are observed with approximately similar
angular and spectral resolutions, ($ii$) the lines are clearly detected ($\ga 5\sigma$), and
($iii$) the line profiles are symmetric.

\subsection{The Milky Way dark cloud L1599B}

In this subsection we consider a source with
narrow [C\,{\sc ii}], [C\,{\sc i}] and CO emission lines observed in the Milky Way~--- the dark cloud L1599B
which is a portion of the molecular ring at $\simeq 27$ pc from the star $\Lambda$ Ori
located at a distance of 425 pc (Goldsmith \etal\ 2016).
Observations of the [C\,{\sc ii}] FS line were obtained with the SOFIA (Stratospheric Observatory for
Infrared Astronomy) airborne observatory, while the [C\,{\sc i}] $J=1$--0, 
$^{12}$CO and $^{13}$CO $J=1$--0 and 2--1, and $^{12}$CO $J=3$--2
data were collected with the Purple Mountain Observatory and 
APEX (Atacama Pathfinder EXperiment) telescopes.
The line profiles shown in Fig.~2 of Goldsmith \etal\ (2016) allow us to select symmetric lines
observed towards two positions O3 and O1 within the L1599B cloud. 
Their local standard of rest (LSR) radial velocities are listed in Table~\ref{T1}.
The $^{13}$CO(2-1) line towards O1 was not detected.
 
Table~\ref{T1} also includes
the [C\,{\sc i}] $^3$P$_{1}$--$^3$P$_{0}$ transition at $\omega = 492160.651(55)$ MHz
(Yamamoto \& Saito 1991).
The corresponding uncertainty $\sigma_v$([C\,{\sc i}]) = 0.034 km~s$^{-1}$ is almost six times lower
than that of [C\,{\sc ii}]
and, thus, the combination of [C\,{\sc ii}]~+~[C\,{\sc i}] imposes the same systematic
error on \daa\ as does the [C\,{\sc ii}] line alone,
i.e. $\sigma_{\rm sys} = 3\times10^{-7}$.
In Table~\ref{T1}, the rest frame frequency of the $^{12}$CO $J = 3 - 2$ transition is
$\omega = 345795.9899(5)$ MHz, and $\omega = 220398.70056(12)$ MHz 
for the strongest among the hyper-fine structure (HFS) lines of
$^{13}$CO $J = 2 - 1$, $F = 5/2 - 3/2$ (Endres \etal\ 2016).
We note that the
three HFS lines of the $^{13}$CO $J = 2 - 1$ transition are split by 0.11 \kms\ between 
the strongest and the weakest ($F = 3/2 - 3/2$) lines which bracket this interval. 
The intensity of the weakest line is about 10 times lower as compared with the 
strongest component\footnote{See, e.g., http://www.splatalogue.net/},
whereas the middle line with $F = 3/2 - 1/2$ and half the intensity of the strongest component
is split from the latter by only 0.045 \kms.
These HFS lines are not resolved in 
observations of the O3 and O1 regions since their widths (FWHM) are larger than 0.5 \kms.

According to Goldsmith \etal\ (2016), the listed emission lines were observed with the following
angular resolutions (FWHM): $\theta$([C\,{\sc ii}])~= $15''$, $\theta$([C\,{\sc i}])~= $12.4''$,
$\theta$($^{12}$CO$_{3-2}$)~= $17.5''$, and 
$\theta$($^{12}$CO$_{2-1}$)~= $\theta$($^{13}$CO$_{2-1}$)~= $27''$--$28''$.
The $^{12}$CO(1-0) line was not included in Table~\ref{T1}
since it was observed with a beam size $\theta = 50''$
albeit its radial velocity is consistent with the other data.

At first, for each dataset O3 and O1, 
we average the radial velocities of CO transitions and compare their
mean weighted values with $V_{\rm [C\,II]}$. The results obtained 
by averaging with weights inversely proportional to 
$\sigma^2_{\scriptscriptstyle {\Delta V}}$ are 
$\langle V_{\rm CO} \rangle = 8.75\pm0.04$ \kms\ (O3), and $\langle V_{\rm CO} \rangle = 11.40\pm0.01$ \kms\ (O1).
This leads to the following constraints on variations in $\alpha$:
$\Delta \alpha/\alpha = (1 \pm 2_{\rm stat} \pm 3_{\rm sys})\times10^{-7}$ (O3), and
$\Delta \alpha/\alpha = (7 \pm 2_{\rm stat} \pm 3_{\rm sys})\times10^{-7}$ (O1).

Then, we calculate the mean radial velocities for the combination [C\,{\sc ii}] + [C\,{\sc i}]:
$\langle V_{\rm [C\,II]+[C\,I]}\rangle = 8.889\pm0.016$ \kms\ (O3), and
$\langle V_{\rm [C\,II]+[C\,I]}\rangle = 11.25\pm0.07$ \kms\ (O1).
This combination gives slightly lower random errors:
$\Delta \alpha/\alpha = (-2.3 \pm 0.7_{\rm stat} \pm 3.0_{\rm sys})\times10^{-7}$ (O3), and
$\Delta \alpha/\alpha = (2.5 \pm 1.2_{\rm stat} \pm 3.0_{\rm sys})\times10^{-7}$ (O1).
Both results are consistent with zero fractional changes in $\alpha$ at the level 
restricted by the laboratory error of the [C\,{\sc ii}] FS transition.

The same order of magnitude bound on $\alpha$ variation 
can be obtained from the analysis of the five Galactic molecular clouds
where the mean velocity offset 
$\langle \Delta V \rangle = 0.022 \pm 0.082_{\rm stat} \pm 0.034_{\rm sys}$ \kms\
between the [C\,{\sc i}] $J = 1-0$ and 
$^{13}$CO $J = 1-0$, $J = 2-1$ transitions
was found in Levshakov, Molaro \& Reimers (2010). 
When interpreted in terms of $\alpha$ variation, it gives
$\Delta \alpha/\alpha = (0.4 \pm 1.4_{\rm stat} \pm 0.6_{\rm sys})\times10^{-7}$.
The advantage of the combination of the [C\,{\sc i}] and CO transitions is in a lower
systematic error. 
Thus, \daa\ is restricted to a few $10^{-7}$ 
in the Milky Way and, as we will see below, also in the nearby galaxy M33.

\begin{table}
\centering
\caption{Radial velocities $V_{\scriptscriptstyle\rm LSR}$
(in \kms) and their $1\sigma$ errors 
measured at the selected positions O3 and O1 towards L1599B
(Goldsmith \etal\ 2016). }
\label{T1}
\begin{tabular}{l r@{$\pm$}l r@{$\pm$}l c} 
\hline
\multicolumn{1}{c}{Spectral Line} & \multicolumn{2}{c}{O3} & \multicolumn{2}{c}{O1} & Beam size\\[-2pt]
 & \multicolumn{2}{c}{ }  &\multicolumn{2}{c}{ } &{\tiny (FWHM)}, \\[-2pt]
 & \multicolumn{2}{c}{ }  &\multicolumn{2}{c}{ }  &  arcsec \\
\hline
[C\,{\sc ii}] $^2$P$_{3/2}$--$^2$P$_{1/2}$ & 8.70&0.12 &  11.0&0.1 & 15\\[2pt]
$^{12}$CO $J$=2--1                         & 8.93&0.03 & 11.39&0.01 & 27-28\\
$^{12}$CO $J$=3--2                         & 8.73&0.01 & 11.41&0.01 & 17.5\\
$^{13}$CO $J$=2--1                         & 8.76&0.02 & \multicolumn{2}{c}{$-$} & 27-28\\[2pt]
[C\,{\sc i}] $^3$P$_{1}$--$^3$P$_{0}$      & 8.89&0.01 & 11.27&0.03 & 12.4\\ 
\hline
\end{tabular}
\end{table}

\subsection{The galaxy M33}

For the recent years, the {\it Herschel} observatory explored 
a large number of nearby galaxies in the FS [C\,{\sc ii}] 158 $\mu$m line.
In this section, we use results of the precise measurements of [C\,{\sc ii}] emission in the 
nearby Triangulum galaxy M33 located at a distance of about $D \sim 800$ kpc 
(McConnachie \etal\ 2009; Brunthaler \etal\ 2005),
namely, three H\,{\sc ii} regions 
located either near the centre of M33 or a few kpc to the north.

The CO(2-1) emission towards the disk of M33 was mapped 
at the 30m IRAM (Institute for Radio Astronomy in the Millimeter Range)
telescope with a resolution $12''\times 2.6$ \kms\ (Gratier \etal\ 2010). 
The  H\,{\sc ii} regions BCLMP691 and BCLMP302 (3.3 kpc and 2 kpc to the north of 
the dynamical centre of M33, respectively)
and the centre of M33 itself were mapped in the [C\,{\sc ii}] line  
with the HIFI (Heterodyne Instrument for the Far Infrared) 
spectrometer onboard the {\it Herschel} satellite
with the same $12''$ angular resolution and 1.2 \kms\ spectral resolution after smoothing 
(Mookerjea \etal\ 2016; Braine \etal\ 2012).

\begin{figure}
 \includegraphics[width=\columnwidth]{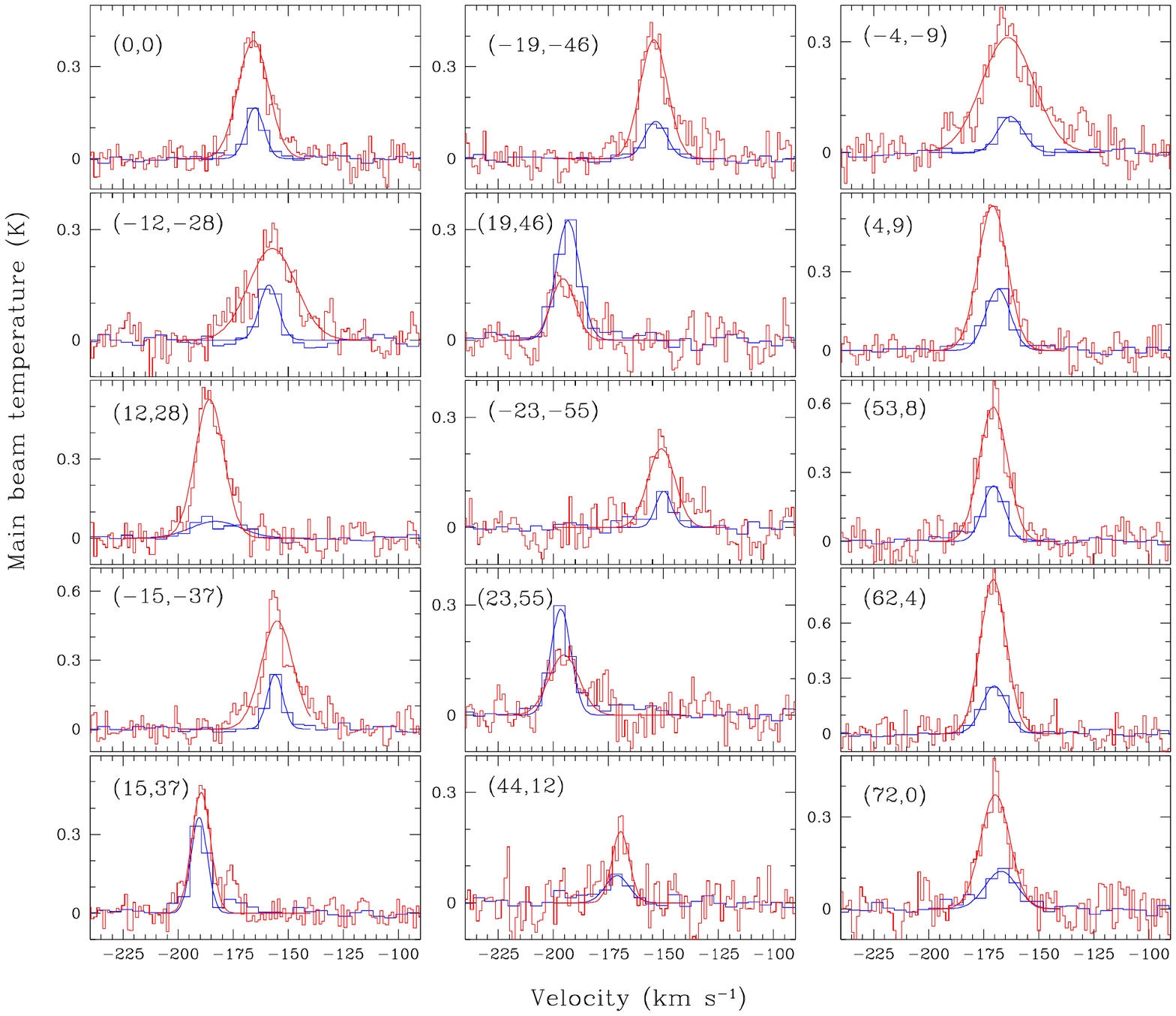}
 \caption{Shown by histograms are the [C\,{\sc ii}] (red) and CO(2-1) (blue) 
spectra at selected positions in the centre of M33.
The one-component Gaussians of the corresponding colours are the fits to the observed spectra.  
The selected positions in arcsec are indicated in parentheses. The same positions are listed in
Table~\ref{T2}. }
 \label{Fg1}
\end{figure}

\begin{figure}
 \includegraphics[width=\columnwidth]{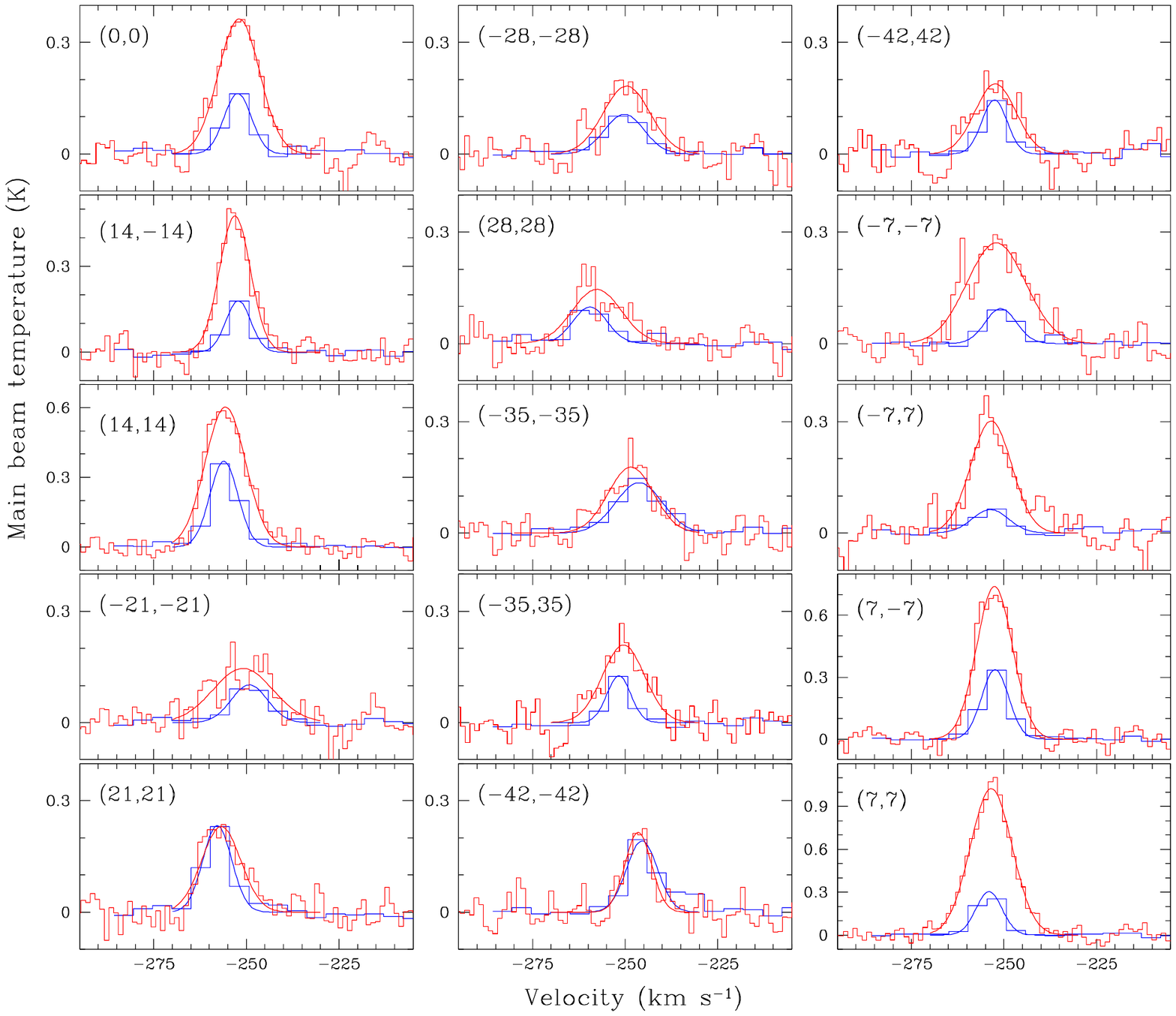}
 \caption{Same as Fig.~\ref{Fg1}, but for selected positions in the BCLMP302 region.}
 \label{Fg2}
\end{figure}

Table~\ref{T2} gives the positions of the spectra (first two columns) with respect to the (0,0) 
offsets which are (J2000) RA = 01:34:16.40, Dec = 30:51:54.6 for the BCLMP691 region 
(Braine \etal\ 2012),
RA = 01:33:48.20, Dec = 30:39:21.4 for the central region, and
RA = 01:34:06.30, Dec = 30:47:25.3 for the BCLMP302 region, respectively (Mookerjea \etal\ 2016).
Columns 3 and 4 give the LSR radial velocities 
and their uncertainties indicated in parentheses which are results of 
one-component Gaussian fits provided by 
the CLASS fitting routine\footnote{http://www.iram.fr/IRAMFR/GILDAS}.
The data for the BCLMP691 region are taken from Table~1 in Braine \etal\ (2012),
whereas the line positions and their errors for the central region and BCLMP302
are calculated in the present paper and the one-component Gaussian fits 
are shown in Figs.~\ref{Fg1} and \ref{Fg2}. 
The velocity offsets $\Delta V = V_{\rm rot} - V_{\rm fs}$ 
are given in the last column 5. 

In total, Table~\ref{T2} lists 46 positions where velocity offsets are known with different uncertainties,
$\sigma_{\scriptscriptstyle {\Delta V}}$.
Being averaged with weights inversely proportional to 
$\sigma^2_{\scriptscriptstyle {\Delta V}}$, the mean offset is equal to
$\langle \Delta V \rangle = -0.01\pm0.14$ \kms.
Using Eq.(\ref{Eq2}), one gets 
$\Delta\alpha/\alpha = (-0.1 \pm 2.4_{\rm stat} \pm 3.0_{\rm sys})\times10^{-7}$.
The result obtained shows
no changes in $\alpha$ at the sensitivity level which is determined by the 
currently available
precision of the laboratory frequencies.

\begin{table}
\centering
\caption{Selected positions observed in [C\,{\sc ii}] and CO(2-1) emission
towards M33.
Given in parentheses are statistical errors $(1\sigma)$.}
\label{T2}
\begin{tabular}{r@{.}l r@{.}l c c r} 
\hline
\multicolumn{2}{c}{$x('')$} & \multicolumn{2}{c}{$y('')$} & $V_{\rm [C\,II]}$ & 
$V_{\rm CO}$ & \multicolumn{1}{c}{$\Delta V$} \\
\multicolumn{2}{c}{ } & \multicolumn{2}{c}{ } & km~s$^{-1}$ & km~s$^{-1}$ & km~s$^{-1}$ \\
\hline
\multicolumn{7}{c}{BCLMP691}\\
$-1$&9 & $-4$&6 & $-269.6(0.2)$ &  $-269.5(0.2)$  &  0.1(0.3)\\[-2pt]
6&5    &   2&7  & $-268.9(0.2)$ &  $-268.9(0.2)$  &  0.0(0.3)\\[-2pt]
4&6    & $-1$&9 & $-269.0(0.3)$ &  $-269.0(0.2)$  &  0.0(0.4)\\[-2pt]
$-2$&7 &   6&5  & $-268.8(0.2)$ &  $-268.6(0.2)$  &  0.2(0.3)\\[-2pt]
$-3$&8 &  $-9$&2& $-268.9(0.6)$ &  $-269.0(0.3)$  &  $-0.1(0.7)$\\[-2pt]
3&8    &   9&2  & $-269.2(0.7)$ &  $-269.1(0.5)$  &  0.1(0.9)\\[-2pt]
$-6$&5 & $-2$&7 & $-269.3(0.2)$ &  $-269.3(0.4)$  &  0.0(0.4)\\[-2pt]
$9$&2 & $-3$&8  & $-265.1(0.4)$ &  $-266.9(0.1)$  & $-1.8(0.4)$\\[-2pt]
$-9$&2 & 3&8    & $-268.3(0.5)$ &  $-268.2(0.7)$  &  0.1(0.9)\\[-2pt]
2&7    & $-6$&5 & $-269.4(0.2)$ &  $-269.4(0.2)$  &  0.0(0.3)\\[-2pt]
1&9    & 4&6    & $-269.1(0.2)$ &  $-269.3(0.2)$  &  $-0.2(0.3)$\\[-2pt]
13&9  & $-5$&7  & $-265.1(0.4)$ &  $-266.1(0.2)$  &  $-1.0(0.4)$\\[-2pt]
$-4$&6 & 1&9   &  $-269.1(0.2)$ &  $-269.4(0.2)$  &  $-0.3(0.3)$\\[-2pt]
9&6   & 23&1  &   $-267.5(0.7)$ &  $-267.2(0.4)$  &   0.3(0.8)\\[-2pt]
18&5  & $-7$&7 &  $-266.7(0.7)$ &  $-265.9(0.4)$  &   0.8(0.8)\\[-2pt]
0&0   &  0&0   &  $-269.4(0.1)$ &  $-269.9(0.2)$  &  $-0.5(0.2)$\\
\multicolumn{7}{c}{M33 centre}\\
0&0     &   0&0   & $-165.8(0.2)$ & $-165.1(0.3)$ &   0.7(0.4)\\[-2pt]
$-12$&0 & $-28$&0 & $-157.2(0.6)$ & $-158.8(0.3)$ & $-1.6(0.7)$\\[-2pt]  
$12$&0   & $28$&0  & $-185.8(0.2)$ & $-182.9(0.8)$ & $2.9(0.8)$\\[-2pt]  
$-15$&0  & $-37$&0 & $-155.0(0.3)$ & $-156.0(0.2)$ & $-1.0(0.4)$\\[-2pt]  
$15$&0   & $37$&0  & $-189.6(0.2)$ & $-190.5(0.1)$ & $-0.9(0.2)$\\[-2pt]  
$-19$&0  & $-46$&0 & $-154.4(0.2)$ & $-153.5(0.3)$ & $0.9(0.4)$\\[-2pt]  
$19$&0   & $46$&0  & $-195.6(0.4)$ & $-193.4(0.2)$ & $2.2(0.4)$\\[-2pt]  
$-23$&0 & $-55$&0 & $-150.9(0.5)$ & $-149.9(0.3)$ & $1.0(0.6)$\\[-2pt]
$23$&0   & $55$&0  & $-195.2(0.7)$ & $-196.6(0.2)$ & $-1.4(0.7)$\\[-2pt]  
$44$&0   & $12$&0  & $-169.4(0.5)$ & $-171.0(0.9)$ & $-1.6(1.0)$\\[-2pt]  
$-4$&0  & $-9$&0  & $-164.0(0.6)$ & $-162.8(0.5)$ &   1.2(0.8)\\[-2pt]
$4$&0   & $9$&0   & $-170.9(0.2)$ & $-168.5(0.1)$ & 2.4(0.2)\\[-2pt]
$53$&0  & $8$&0   & $-170.6(0.3)$ & $-170.5(0.2)$ & 0.1(0.4)\\[-2pt]
62&0    &  4&0    & $-170.6(0.2)$ & $-170.1(0.3)$ &   0.5(0.4)\\[-2pt]
72&0    &  0&0    & $-169.7(0.5)$ & $-167.0(0.7)$ &   2.7(0.9)\\
\multicolumn{7}{c}{BCLMP302}\\
0&0     &  0&0    & $-252.0(0.2)$ & $-252.3(0.4)$ &   $-0.3(0.4)$\\[-2pt]
14&0    & $-14$&0 & $-253.1(0.2)$ & $-252.2(0.4)$ &       0.9(0.4)\\[-2pt]
14&0    &  14&0   & $-255.7(0.2)$ & $-256.1(0.1)$ &   $-0.4(0.2)$\\[-2pt]
$-21$&0 & $-21$&0 & $-250.9(0.7)$ & $-249.2(0.5)$ &  1.7(0.9)\\[-2pt]
$21$&0  & $21$&0  & $-256.8(0.3)$ & $-257.9(0.3)$ &  $-1.1(0.4)$\\[-2pt]
$-28$&0 & $-28$&0 & $-249.6(0.7)$ & $-250.2(0.4)$ &  $-0.6(0.8)$\\[-2pt]
$28$&0  & $28$&0  & $-257.7(0.6)$ & $-259.5(0.5)$ &  $-1.8(0.8)$\\[-2pt]
$-35$&0 & $-35$&0 & $-248.5(0.5)$ & $-246.3(0.4)$ &  $2.2(0.6)$\\[-2pt]
$-35$&0 & $35$&0  & $-250.5(0.5)$ & $-251.7(0.5)$ &  $-1.2(0.7)$\\[-2pt]
$-42$&0 & $-42$&0 & $-246.5(0.4)$ & $-245.5(0.2)$ &  $1.0(0.4)$\\[-2pt]
$-42$&0 & $42$&0  & $-252.3(0.6)$ & $-252.6(0.2)$ &  $-0.3(0.6)$\\[-2pt]
$-7$&0  & $-7$&0  & $-252.1(0.5)$ & $-250.9(0.7)$ &  $1.2(0.9)$\\[-2pt]
$-7$&0  &  7&0    & $-253.5(0.3)$ & $-253.7(0.7)$ &  $-0.2(0.8)$\\[-2pt]
7&0     & $-7$&0  & $-252.6(0.2)$ & $-252.4(0.2)$ &     0.2(0.3)\\[-2pt]
7&0     &  7&0    & $-253.5(0.1)$ & $-254.1(0.1)$ &   $-0.6(0.1)$\\
\hline
\end{tabular}
\end{table}

\subsection{High-redshift [C\,{\sc ii}]/CO systems}

Here we take up a sample of five $z > 5.7$ quasars with detected  
[C\,{\sc ii}] FS and CO rotational transitions.
As mentioned in Sect.~1, the electron-to-proton mass ratio is stable to
an accuracy of a few $10^{-6}$ up to redshift $z \sim 4$.
In this case, the FST method can be applied for estimating changes in $\alpha$
within the interval 
$10^{-5} \la |\Delta\alpha/\alpha| \la 10^{-4}$.

At present, the accuracy of the high-$z$ estimates of \daa\ by the FST method
is about two orders of magnitude
lower as compared with the local Universe because high-redshift objects are 
fainter, signal-to-noise ratios are lower, and spectral resolution is moderate.
Nevertheless, this method has a perspective since it enables us to test $\alpha$
at cosmological epochs which are not accessible to optical spectroscopy.

The selected sources showing symmetric emission line profiles are listed in Table~\ref{T3}.
Column 1 provides the quasar names, columns 2 and 3 give the measured redshifts
of the [C\,{\sc ii}] 158 $\mu$m and CO lines as described below, and the velocity offsets
$\Delta V = V_{\rm CO} - V_{\rm [C\,II]} = 
c(z_{\rm CO} - z_{\rm [C\,II]})/(1 + \bar{z})$ are presented in column~4 
(here $\bar{z}$ is the mean redshift).

{\it J0129--0035.} 
The [C\,{\sc ii}] line was observed at $z = 5.8$ with the
Atacama Large Millimeter/submillimeter Array (ALMA) 
at $0.57'' \times 0.49''$ resolution (Fig.~5 in Wang \etal\ 2013),
while the CO(6-5) line\footnote{The rest frame frequency is $\omega = 691473.0763(5)$ MHz (Endres \etal\ 2016).}
was observed at 3~mm with the IRAM Plateau de Bure Interferometer (PdBI)
at $\sim 5''$ resolution (Fig.~2 in Wang \etal\ 2011).
Table~\ref{T3} gives the host galaxy redshifts measured with the [C\,{\sc ii}] and CO(6-5) lines.

{\it J2310+1855.}
Both the [C\,{\sc ii}] and CO(6-5) lines were detected at $z = 6.0$ by Wang \etal\ (2013).
[C\,{\sc ii}] was observed with ALMA at $0.72'' \times 0.51''$ resolution,
while
the CO(6-5) line was observed with the PdBI at the beam size $5.4'' \times 3.9''$ 
(Figs.~1 and 5 in Wang \etal\ 2013).
The measured redshifts of these lines are listed in Table~\ref{T3}.

{\it J2054--0005.}
The [C II] emission line was observed at $z = 6.0$ 
with ALMA at $0.64'' \times 0.58''$ resolution (Fig.~5 in Wang \etal\ 2013), 
while CO(6-5) was detected with the PdBI at $\sim 5''$ resolution (Fig.~1 in Wang \etal\ 2010). 
The redshifts of these lines are given in Table~\ref{T3}.

{\it J1319+0950.}
The [C\,{\sc ii}] emission line was observed at $z = 6.1$ 
with a synthesized beam size of $0.69'' \times 0.49''$ using ALMA (Fig.~5 in Wang \etal\ 2013). 
Wang \etal\ (2011) show in their Fig.~1 the observed profile of the CO(6-5) line which was obtained with
the 3 mm receiver on the PdBI at $\sim 3.5''$ resolution
The measured redshifts are presented in Table~\ref{T3}.

{\it J1148+5251.}  
The CO(7-6)\footnote{The rest frame frequency is $\omega = 806651.806(5)$ MHz (Endres \etal\ 2016).} 
and (6-5) emission lines were detected at $z = 6.4$ by Bertoldi \etal\ (2003)
using the PdBI at $\sim 5''$ resolution.
The observed spectra of CO(7-6) and (6-5) are shown in their Fig.~1.
The measured redshifts of 
$z_{\rm CO(7-6)} = 6.4192\pm0.0009$ and $z_{\rm CO(6-5)} = 6.4187\pm0.0006$ yield the mean value
$\langle z_{\rm CO}\rangle = 6.4190\pm0.0005$ which is listed in Table~\ref{T3}.
The [C\,{\sc ii}] line emission
redshifted to $z = 6.4189\pm0.0006$ was observed by Maiolino \etal\ (2005)
with the IRAM~30m telescope at $9.6''$ resolution (see their Fig.~1).
On the other hand, Walter \etal\ (2009) who used the PdBI at $\sim 0.3''$ resolution
determined a central velocity of the [C\,{\sc ii}] line of $3\pm12$ \kms\
relative to the CO redshift of $z=6.419$ (see their Fig.~2), which leads to
$z_{\rm [C\,II]} = 6.4189\pm0.0003$.
Both results yield the mean $\langle z_{\rm [C\,II]}\rangle = 6.4189\pm0.0003$
given in Table~\ref{T3}.

These five velocity offsets $\Delta V$, being 
averaged with weights inversely proportional to their uncertainties squared,
yield $\langle \Delta V \rangle = 0\pm8$ \kms, and thus $|\Delta\alpha/\alpha| < 1.3\times10^{-5}$
at redshift $z = 5.7 - 6.5$.
The result obtained improves by five times our previous constraint on \daa\ at $z = 6.4$ (Levshakov \etal\ 2008).
Thus, we may conclude that $\alpha$ does not vary at the level of $\sim 10^{-5}$ 
up to $z = 6.5$,
corresponding to 5\% of the age of the Universe today. 

To complete this section we note that the [C\,{\sc ii}] 158 $\mu$m emission was observed 
towards a number of distant quasars and galaxies with $z > 6$
where the CO lines have not yet been detected. 
These objects are listed in Table~\ref{T4}.
The redshifts for some of them were also
determined from the Mg\,{\sc ii} $\lambda2798$ \AA\ low ionization line,
and it is tempting to use the measured redshifts $z_{\rm Mg\,II}$ and $z_{\rm [C\,II]}$ to
put constraints on \daa, 
for the ratio of frequencies of the FS to the optical gross-structure transitions 
is proportional to $\alpha$ squared (FST method, see Sect.~1).
However, for the  $z > 6$ objects, it was found that the peaks of the Mg\,{\sc ii}
and [C\,{\sc ii}] or CO lines show significant offsets from each other with
the Mg\,{\sc ii} line being
blueshifted by $\Delta V = -480\pm630$ \kms\ (Venemans \etal\ 2016), 
which inhibits accurate measurements of \daa\ using optical spectra.
We can only hope that
other molecular lines which closely trace the [C\,{\sc ii}] 158 $\mu$m emission
will be detected in the future. Then more stringent limits than 
our $10^{-5}$ value on \daa\
could be obtained by radio methods for this early cosmological epoch. 

\begin{table}
\centering
\caption{Velocity offsets $\Delta V = c(z_{\rm CO} - z_{\rm [C\,II]})/(1 + \bar{z})$ for the $z > 5.7$ quasars.
Given in parentheses are statistical errors $(1\sigma)$.
References: 
(1) Wang \etal\ 2013,
(2) Wang \etal\ 2011,
(3) Wang \etal\ 2010,  
(4) Bertoldi \etal\ 2003, 
(5) Maiolino \etal\ 2005, 
(6) Walter \etal\ 2009. 
}
\label{T3}
\begin{tabular}{l l l r@{$\pm$}l } 
\hline
\multicolumn{1}{c}{Name} & \multicolumn{1}{c}{$z_{\rm [C\,II]}$} & \multicolumn{1}{c}{$z_{\rm CO}$} &
\multicolumn{2}{c}{$\Delta V$, \kms} \\
\hline
J0129--0035       & 5.7787(1)$^1$ &  5.7794(8)$^2$ & 31&35 \\ 
J2310+1855        & 6.0031(2)$^1$ &  6.0025(7)$^1$ & $-30$&30 \\ 
J2054--0005       & 6.0391(1)$^1$ &  6.0379(22)$^3$ & $-50$&90 \\ 
J1319+0950        & 6.1330(7)$^1$ &  6.1321(12)$^2$ & $-40$&60 \\ 
J1148+5251        & 6.4189(3)$^{5,6}$ &  6.4190(5)$^4$ & 4&12 \\ 
\hline
\end{tabular}
\end{table}

\begin{table}
\centering
\caption{[C\,{\sc ii}] measurements for the $z > 6$ quasars and galaxies
without detected CO lines.
Given in parentheses are statistical errors $(1\sigma)$.
References: 
(1) Willott \etal\ 2015a,
(2) Willott \etal\ 2015b,
(3) Jones \etal\ 2017,  
(4) Willott \etal\ 2013, 
(5) Venemans \etal\ 2016, 
(6) Brada$\check{c}$ \etal\ 2017, 
(7) Smit \etal\ 2017,
(8) Venemans \etal\ (2017).
}
\label{T4}
\begin{tabular}{l l l c } 
\hline
\multicolumn{1}{c}{Name} & \multicolumn{1}{c}{$z_{\rm [C\,II]}$} & \multicolumn{1}{c}{Name} &
\multicolumn{1}{c}{$z_{\rm [C\,II]}$} \\
\hline
J0055+0146        & 6.0060(8)$^1$ & RXJ1347:1216 & 6.7655(5)$^6$\\ 
WMH~5             & 6.0695(3)$^{2,3}$ & J0109--3047 & 6.7909(4)$^5$ \\ 
J2229+1457       & 6.1517(5)$^1$ & COS2987030247 & 6.8076(2)$^7$ \\ 
CLM~1            & 6.1657(3)$^2$ & COS3018555981 & 6.8540(3)$^7$ \\ 
J0210--0456      & 6.4323(5)$^{4}$ & J2348--3054 & 6.9018(7)$^5$ \\ 
J0305--3150      & 6.6145(1)$^{5}$ & J1120+0641 & 7.0851(5)$^8$ \\ 
\hline
\end{tabular}
\end{table}

\section{Summary and Conclusions}

In this work, we present an analysis of 
[C\,{\sc i}], [C\,{\sc ii}], and CO lines of
one molecular cloud in the disk of the Milky Way, 
three H\,{\sc ii} zones in the neighbour galaxy M33, 
and five quasar systems at $z = 5.7 - 6.5$.

To obtain a limit on \daa\ by 
the Fine Structure Transition (FST) method (Levshakov \& Kozlov 2017), 
we used recently published observations with the 
{\it Herschel} and SOFIA observatories, the Purple Mountain Observatory, 
APEX, ALMA, and IRAM telescopes.

\begin{figure}
 \includegraphics[width=\columnwidth]{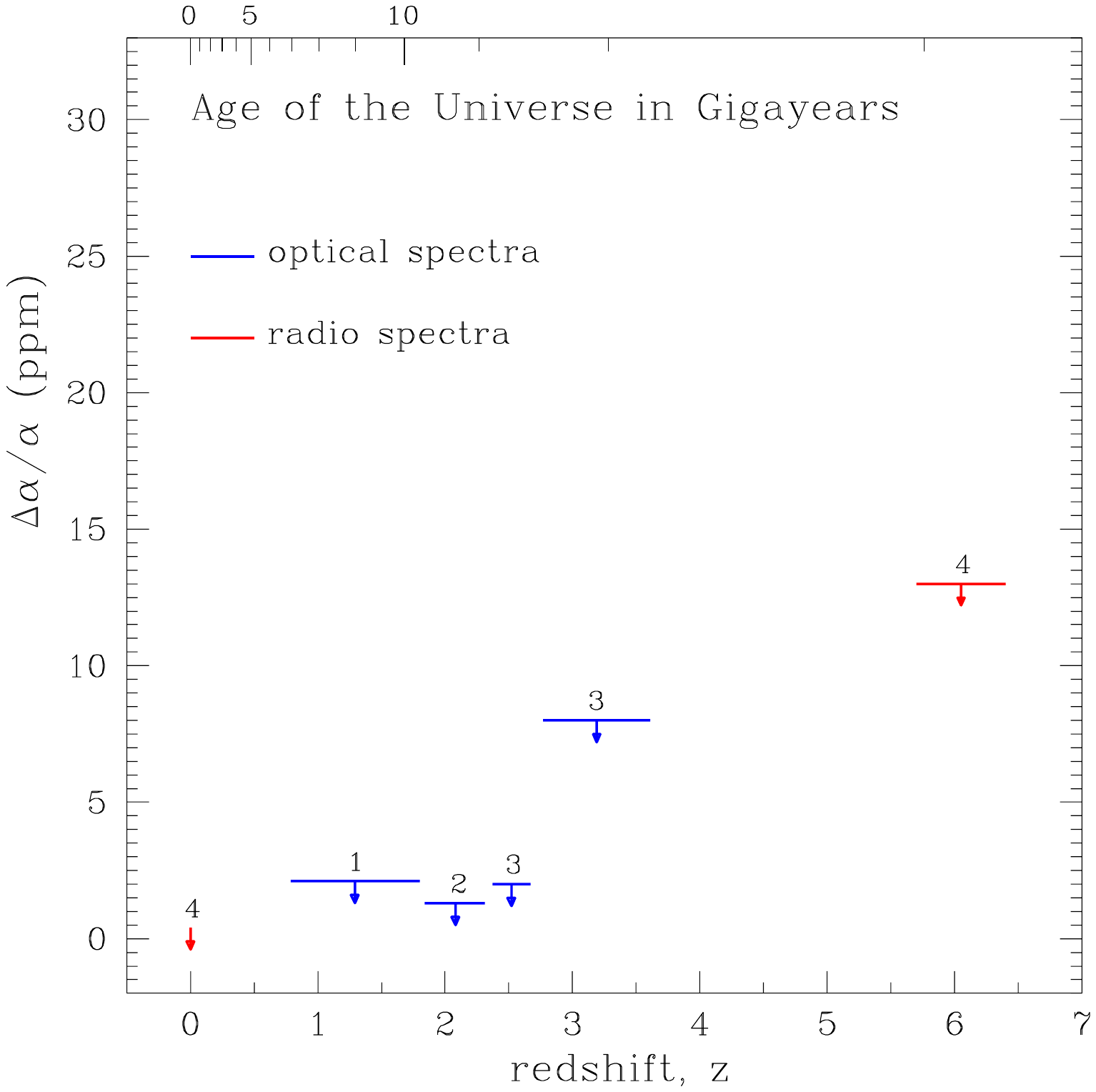}
 \caption{The observational constraints on $\Delta\alpha/\alpha$ in parts per million (ppm)
from optical (blue) and radio (red) observations plotted versus redshift, $z$.
The age of the Universe in Gigayears is shown on the top axis.
The horizontal bars indicate the covered redshift ranges where the $1\sigma$ upper bounds were
calculated. The redshift range of the low redshift radio constraints (Milky Way and M33) is
unresolved at the scale of this plot. The plot illustrates that all available optical constraints
are restricted by $z \la 4$.
References: 1~-- Evans \etal\ 2014; 2~-- Murphy \etal\ 2017; 3~-- King \etal\ 2012; 4~-- this paper.}
 \label{Fg3}
\end{figure}

At low redshifts, the FST method
produces results which agree quantitatively with those obtained by other methods.
However, at high redshifts, it has the distinct advantage over 
optical spectral observations which are applicable
to cosmological tests of \daa\ only up to $z \la 4$.
The comparison between optical and radio constraints on \daa\ based
on the most accurate recent observations is shown in Fig.~\ref{Fg3}.

Our main conclusions are as follows:
\begin{itemize}
\item[$\bullet$]
Using a combination of the [C\,{\sc ii}] and [C\,{\sc i}] fine structure (FS) lines, 
$^{12}$CO(2-1), (3-2), and $^{13}$CO(2-1) being observed towards two 
positions of the dark molecular cloud L1599B in our Galaxy,
we set a limit on $|\Delta\alpha/\alpha| < 3\times10^{-7}$.
This limit is in agreement with $|\Delta\alpha/\alpha| < 2\times10^{-7}$
resulting from the analysis of five Galactic molecular clouds using
[C\,{\sc i}] FS, and $^{13}$CO(2-1), (1-0) lines (Levshakov, Molaro \& Reimers 2010).
\item[$\bullet$]
From the relative positions of the [C\,{\sc ii}] FS and CO(2-1) emission lines 
observed towards three H\,{\sc ii} regions of M33, we find 
the most stringent constraint to date on $\alpha$ variation 
of $|\Delta\alpha/\alpha| < 4\times10^{-7}$ for the neighbour galaxy.
\item[$\bullet$]
The most distant quasar systems from the redshift range $z = 5.7 - 6.5$ 
with detected [C\,{\sc ii}] FS and CO(6-5), (7-6) emission
show no changes in $\alpha$ at the level of
$|\Delta\alpha/\alpha| < 1.3\times10^{-5}$.
\end{itemize}

\section*{Acknowledgements}

We thank our anonymous referee for comments and suggestions that helped
clarifying the content of this paper.

\clearpage

\clearpage

\clearpage

\end{document}